\documentclass[prl,twocolumn,superscriptaddress,showpacs]{revtex4}

\usepackage{graphicx}
\usepackage{dcolumn}
\usepackage{bm}

\begin{document}

\preprint{ATB-1}

\title{Three-Dimensional Spin Fluctuations in Na$_{0.75}$CoO$_2$}

\author{L. M. Helme}
\email{l.helme1@physics.ox.ac.uk} \affiliation{ Department of
Physics, University of Oxford, Oxford, OX1 3PU, United Kingdom }
\author{A. T. Boothroyd}\affiliation{
Department of Physics, University of Oxford, Oxford, OX1 3PU,
United Kingdom }
\author{R. Coldea}\affiliation{
Department of Physics, University of Oxford, Oxford, OX1 3PU,
United Kingdom }
\author{D. Prabhakaran}\affiliation{
Department of Physics, University of Oxford, Oxford, OX1 3PU,
United Kingdom }
\author{D. A. Tennant}\affiliation{
School of Physics and Astronomy, University of St. Andrews, St.
Andrews, Fife KY16 9SS, United Kingdom}
\author{A. Hiess}\affiliation{
Institut Laue-Langevin, BP 156, 38042 Grenoble Cedex 9, France }
\author{J. Kulda}\affiliation{
Institut Laue-Langevin, BP 156, 38042 Grenoble Cedex 9, France }

\date{\today}

\begin{abstract}
We report polarized- and unpolarized-neutron scattering
measurements of magnetic excitations in single-crystal
Na$_{0.75}$CoO$_2$. The data confirm ferromagnetic correlations
within the cobalt-layers and reveal antiferromagnetic correlations
perpendicular to the layers, consistent with an A-type
antiferromagnetic ordering. The magnetic modes propagating
perpendicular to the layers are sharp, and reach a maximum energy
of $\sim$12\,meV. From a minimal spin wave model, containing only
nearest-neighbour Heisenberg exchange interactions, we estimate
the inter- and intra-layer exchange constants to be $12.2 \pm
0.5$\,meV and $-6 \pm 2$\,meV, respectively. We conclude that the
magnetic fluctuations in Na$_{0.75}$CoO$_2$ are highly
three-dimensional.

\end{abstract}

\maketitle


Its potential as a battery electrode material \cite{battery-ref}
and in thermoelectric devices \cite{thermoelectric-ref}, and most
recently the discovery of superconductivity after hydration
\cite{takada-2003}, have made the layered cobaltite
Na$_{x}$CoO$_2$ the subject of intense research in the last few
years. The structure comprises two-dimensional layers of CoO$_2$
separated by layers of sodium ions, with the Co atoms forming a
triangular lattice \cite{structure-ref}. The superconducting
compound Na$_{x}$CoO$_2\cdot y$H$_2$O ($x \approx 0.3$, $y \approx
1.3$, $T_{\rm c} \approx 4.5$\,K) is of particular interest as the
first cobalt-oxide based superconductor \cite{takada-2003}. The
layered structure and existence of superconductivity over a narrow
range of doping \cite{Schaak} near to a Mott insulator invite
comparisons with the copper-oxide superconductors, and evidence is
mounting for an unconventional mechanism of superconductivity
\cite{unconventional}.

In common with the superconducting cuprates, the properties of the
normal metallic state of Na$_{x}$CoO$_2$ are unusual and show
effects due to strong electronic correlations. One of the
outstanding puzzles is the nature of the magnetic interactions,
which may play a central role in the formation of the
superconducting state. In the range $x \sim 0.7$--$0.95$ the
susceptibility of Na$_{x}$CoO$_2$ shows a sharp magnetic
transition at $T_{\rm m} \approx 22$\,K
\cite{motohashi-2003,susceptibility}. Muon-spin rotation ($\mu$SR)
measurements confirmed the existence of static magnetic order
below $T_{\rm m}$, and placed an upper limit of 0.2\,$\mu_{\rm B}$
on the size of the ordered moment \cite{muSR}. For $T
> T_{\rm m}$ the susceptibility can be fitted to a Curie--Weiss law plus a constant
term, which indicates a degree of local character to the magnetism
\cite{motohashi-2003}. Such fits give negative Weiss temperatures,
implying dominant antiferromagnetic correlations. However, a
recent neutron scattering study of Na$_{0.75}$CoO$_2$ found strong
ferromagnetic spin correlations within the cobalt-oxide layers
\cite{boothroyd-may04}, consistent with several theoretical
predictions \cite{In-plane-FM-theory}. Until now there have been
no measurements of magnetic fluctuations perpendicular to the
layers, and the nature of the magnetic order below $T_{\rm m}$ has
not been established.

Here we investigate further the magnetic correlations in
Na$_{0.75}$CoO$_2$, using both polarized- and unpolarized-neutron
scattering. The new data reveal strong antiferromagnetic
correlations perpendicular to the cobalt-oxide layers, consistent
with an A-type antiferromagnetic ordering.  Surprisingly, the
inter-layer exchange coupling is found to be similar in strength
to the intra-layer coupling, so that despite the layered structure
the magnetic interactions are highly three-dimensional.

The measurements were made on a single crystal of
Na$_{0.75}$CoO$_2$ grown in Oxford by the floating-zone method
\cite{prabhaks-paper}. For the neutron studies we cleaved a
crystal of mass $\sim$1.5\,g from a zone-melted rod. Smaller
crystals from the same rod were examined by x-ray diffraction,
magnetometry and electron microscopy. The analysis revealed the
presence of small inclusions of cobalt oxides (CoO and
Co$_3$O$_4$) consistent with previous reports for melt-grown
crystals \cite{Chou}. These impurity phases, which amounted to a
few per cent of the total, were found by neutron diffraction to
grow epitaxially on the host lattice. Once the orientation of the
impurity crystallites had been determined it was straightforward
to distinguish the impurity signal from that of the host. The
anisotropic magnetic susceptibility of the crystals exhibited an
anomaly at $T_{\rm m} \approx 22$\,K for fields applied parallel
to the $c$ axis, consistent with the magnetic transition observed
previously \cite{motohashi-2003}.

Unpolarized- and polarized-neutron scattering measurements were
performed on the thermal triple-axis spectrometers IN8 and IN20,
respectively, at the Institut Laue-Langevin. On IN8 we employed a
Si $(111)$ monochromator and a pyrolytic graphite $(002)$
analyser, and worked with a fixed final energy $E_{\rm f}$ =
14.7\,meV. To increase the count rate both monochromator and
analyser were curved horizontally and vertically for optimum
focussing.  For the polarized-neutron measurements on IN20 we used
curved Heusler $(111)$ as both monochromator and analyser, and
$E_{\rm f}$ = 34.8\,meV. On both instruments a graphite filter was
placed in the scattered beam to suppress higher order harmonics.
The crystal was mounted in a standard helium cryostat, and most of
the measurements were made within the $(100)$--$(001)$ scattering
plane. Here, the notation $(hkl)$ represents the wavevector ${\bf
Q}=h{\bf a}^{\star} + k{\bf b}^{\star} + l{\bf c}^{\star}$, where
${\bf a}^{\star}$, ${\bf b}^{\star}$ and ${\bf c}^{\star}$ are the
basis vectors of the hexagonal reciprocal lattice (the angle
between ${\bf a}^{\star}$ and ${\bf b}^{\star}$ is 60\,deg)
\cite{basis-vectors}.







\begin{figure}
\includegraphics*{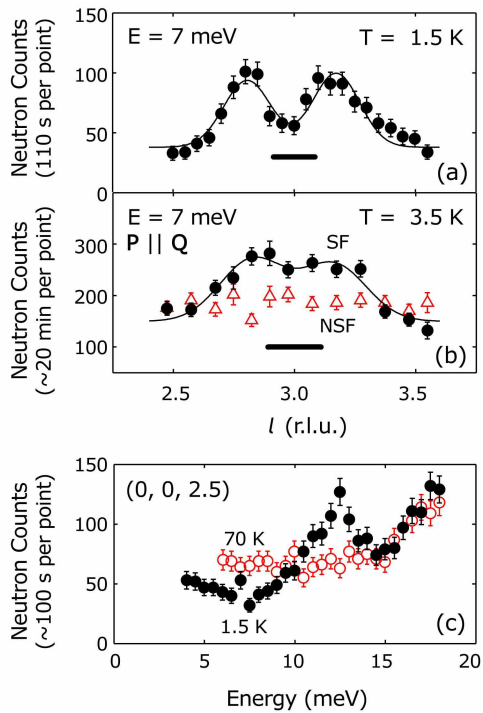}|
\caption{\small (Color online) Magnetic scattering from
Na$_{0.75}$CoO$_2$. (a) and (b) show scans parallel to $(00l)$
made at a fixed energy of 7\,meV. (c) displays energy scans made
at the zone boundary $l=2.5$ at temperatures of 1.5\,K and 70\,K.
The data in (a) and (c) were obtained with unpolarized neutrons on
IN8 using a fixed final energy of $E_{\rm f}$ = 14.7\,meV. The
data in (b) are the spin-flip (SF) and non-spin-flip (NSF)
scattering collected on IN20 with the neutron polarization {\bf P}
parallel to the scattering vector {\bf Q}, and with fixed $E_{\rm
f}$ = 34.8\,meV. The horizontal bars in (a) and (b) indicate the
experimental resolution.} \label{fig:fig1}
\end{figure}

Our previous studies \cite{boothroyd-may04} showed that the
magnetic excitation spectrum of Na$_{0.75}$CoO$_2$ is centred on
the $\Gamma$-point of the two-dimensional Brillouin zone, i.e.\
$(00)$, corresponding to ferromagnetic correlations within the
cobalt-oxide layers. In the present work we concentrated on the out-of-plane wavevector component of the
magnetic fluctuations.

Figure \ref{fig:fig1}(a) shows an example scan parallel to the
$(00l)$ direction performed on IN8 at a fixed energy transfer of
7\,meV. Two peaks can be seen symmetrically either side of $l=3$.
Figure \ref{fig:fig1}(b) displays the same scan but this time
performed on IN20 with the neutron polarization maintained
parallel to the scattering vector during the scan. In this
configuration the spin-flip (SF) scattering is purely magnetic,
and the non-spin-flip scattering is non-magnetic. The two peaks
are clearly present in the SF channel and absent from the NSF
channel. The peaks are essentially resolution-limited, as
indicated, but are less well resolved in Fig. \ref{fig:fig1}(b)
than in Fig. \ref{fig:fig1}(a) because of the larger neutron
energy used on IN20. We conclude that the peaks arise from
magnetic excitations.


The scan shown in Fig. \ref{fig:fig1}(a) was repeated for
different fixed energies between 3~\,meV and 10\,meV. Each scan
contained two peaks symmetric about $(003)$, with the peak
separation increasing with increasing energy. In addition, energy
scans were made at several fixed points along the line $(00l)$).
Figure\ \ref{fig:fig1}(c) shows one such scan, made at $l=2.5$,
the zone boundary in the out-of-plane direction. The scan was
performed at 1.5\,K and then repeated at 70~K.
The prominent peak at $\sim$12\,meV in the low temperature scan
has disappeared by 70\,K. This again confirms the magnetic origin
of the scattering since magnetic correlations are destroyed with
increasing temperature.


\begin{figure}
\includegraphics*{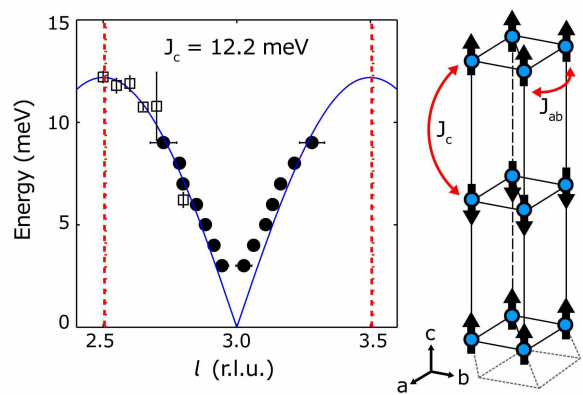}|
\caption{\small (Color online) Left: The magnon dispersion of
Na$_{0.75}$CoO$_2$ parallel to $(00l)$ measured on IN8 at a
temperature of 1.5\,K. Filled circles are from constant-energy
scans, e.g. Fig.\ \ref{fig:fig1}(a), and open squares are from
constant-$l$ scans, e.g. Fig.\ \ref{fig:fig1}(c). The solid curve
is calculated from the spin wave dispersion Eqs.\ \ref{eq:2} and
\ref{eq:3} with exchange constant $J_c = 12.2$\,meV. Dotted lines
show the zone boundaries. Right: The magnetic structure on which
the spin wave model is based, showing the two exchange constants
$J_{ab}$ and $J_c$, and the spin direction. \label{fig:fig2}}
\end{figure}

By fitting Gaussian functions to the peaks in both types of scan
we constructed the magnon dispersion relation. This is displayed
in Fig.\ \ref{fig:fig2}. There is clearly a mode dispersing from
$(003)$ with a maximum energy of approximately 12\,meV. The
crystal structure of Na$_{x}$CoO$_2$ is such that no structural
Bragg peaks are allowed for positions $(00l)$ with odd $l$. As
expected, therefore, no structural Bragg peak was observed at
$(003)$, but scans made at different temperatures revealed no
magnetic Bragg peak at this point either.


The simplest spin arrangement consistent with the observations is
the A-type antiferromagnet shown in Fig.\ \ref{fig:fig2}, in which
the spins are ordered ferromagnetically within the layers and the
layers are coupled antiferromagnetically along the $c$ axis.  Each
cobalt ion is taken to have the same spin. For no magnetic Bragg
peak to appear at $(003)$ the spins must be parallel or
antiparallel to the $c$-axis. This is because neutrons do not
couple to spin components parallel to the scattering vector.

To analyze the three-dimensional dispersion in more detail we
compare the experimental results with a spin wave model containing
the minimum number of exchange parameters. The Hamiltonian is


\begin{equation}\label{Hamiltonian}
    \mathcal{H} = J_{ab}\sum_{\langle i, i^\prime \rangle} {\bf S}_i
    \cdot {\bf S}_{i^\prime} + J_c\sum_{\langle i, j \rangle}  {\bf S}_i
    \cdot {\bf S}_j ,\label{eq:1}
\end{equation}
where $J_{ab}$ and $J_c$ are intra- and inter-layer exchange
parameters, respectively, as indicated in Fig.\ \ref{fig:fig2}.
Only nearest-neighbour interactions are included in the
summations, and $\langle i, i^\prime \rangle$ and $\langle i, j
\rangle$ denote spin pairs within the same layer and on adjacent
layers, respectively.

Standard methods were used to derive the spin-wave dispersion and
scattering cross-section from the Hamiltonian. The resulting
expression for the spin-wave energy dispersion is
\begin{equation}\label{dispersion-relation}
\hbar\omega_{\mathbf k} = 2S\sqrt{(J_{\mathbf k} - J_{{\mathbf
k}_m})[(J_{\mathbf{k} - {\mathbf k}_m}+J_{\mathbf{k} + {\mathbf
k}_m})/2 - J_{{\mathbf k}_m}]},\label{eq:2}
\end{equation}
where $S$ is the spin (here assumed to be $S$ = 1/2), ${\mathbf
k}$ is the magnon wavevector, and ${\mathbf k}_m = (001)$ is the
propagation vector of the magnetic structure. $J_{\mathbf k}$ is
the Fourier transform of the exchange couplings, given by
\begin{equation}\label{J_k}
    J_{\mathbf k} = J_c \cos(\pi l) + J_{ab}[\cos(2\pi h) + \cos(2\pi k) + \cos(2\pi (h + k))]
    ,\label{eq:3}
\end{equation}
with $J_{{\mathbf k}_m}$, $J_{\mathbf{k} - {\mathbf k}_m}$ and
$J_{\mathbf{k} + {\mathbf k}_m}$ defined in a similar manner.

The dispersion relation along the $(00l)$ direction does not
depend on $J_{ab}$, so by comparing the spin-wave dispersion to
the data in Fig.\ \ref{fig:fig2} we can immediately obtain a value
for $J_c$. The best fit is shown by the solid curve on Fig.\
\ref{fig:fig2}, which is calculated with $J_c=12.2$\,meV. At low
energies the data points lie systematically above the fitted
curve, suggesting the presence of a small gap of 1--2\,meV. Apart
from this, the model provides a good description of the data.


\begin{figure}
\includegraphics*{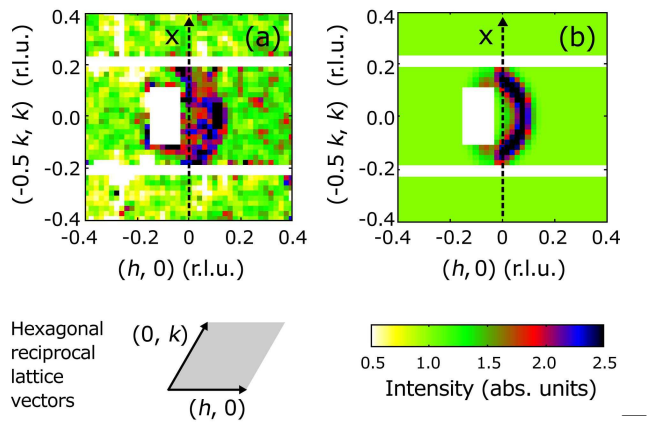}|
\caption{\small (Color online) (a) Neutron inelastic scattering
from Na$_{0.75}$CoO$_2$ recorded on the MAPS spectrometer at ISIS
with an incident energy of 60\,meV \cite{boothroyd-may04}. The map
contains data averaged over energy transfers of 8--12\,meV, and is
projected onto the $(h,k)$ reciprocal lattice plane of the
crystal. (b) Simulated intensity using the model described in the
text with $J_{ab}=-6$\,meV and $J_c=12.2$\,meV. The axis labels
correspond to the hexagonal reciprocal axes drawn in the figure.}
\label{fig:fig3}
\end{figure}

\begin{figure}
\includegraphics*{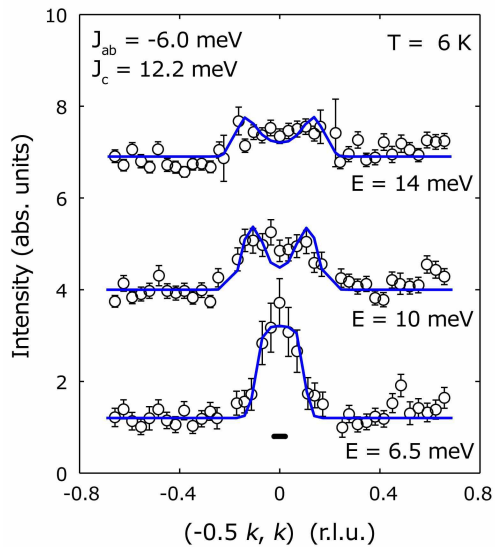}|
\caption{\small (Color online) Constant energy cuts taken along
the line marked X in Fig.\ \ref{fig:fig3}. Open circles show
neutron data points, while the solid lines are from the
simulation. The 10\,meV and 14\,meV data have been displaced
vertically by 3 and 6 units, respectively. The horizontal bar
indicates the resolution.} \label{fig:fig4}
\end{figure}

The measurements described so far probe only the inter-plane
correlations. To gain quantitative information on correlations
within the planes we apply the model to the results obtained
previously with a crystal of the same composition using the MAPS
time-of-flight spectrometer at ISIS \cite{boothroyd-may04}. Figure
\ref{fig:fig3}(a) reproduces part of a slice through the MAPS data
from Ref. \cite{boothroyd-may04}, in which the data were averaged
over energy transfers $E=8-12$\,meV. The configuration used on
MAPS is such that the out-of-plane wavevector component $l$ varies
with $E$. For this slice $l\approx 1$.

To continue the analysis, a MAPS-style data set was simulated from
the model to allow direct comparison with the MAPS data. The MAPS
data set is an intensity array in ({\bf Q}, {\it E}) space, so for
each data point in this space the simulated intensity was
calculated, including the magnetic form-factor and orientation
factor \cite{Squires}. For the calculation, $J_c$ was fixed to the
value 12.2\,meV determined from the inter-layer dispersion, while
$J_{ab}$ was varied until good agreement between simulation and
experiment was achieved.

Following this procedure we determined that $J_{ab} = -6\pm
2$\,meV.  Figure \ref{fig:fig3}(b) shows a slice through the
simulated data for $J_{ab} = -6$\,meV and $J_c=12.2$\,meV to give
direct comparison with Fig.\ \ref{fig:fig3}(a). The distribution
of scattering within the plane is well reproduced by the model.
Figure \ref{fig:fig4} shows constant-energy cuts through both real
and simulated data sets along the line marked X in Fig.\
\ref{fig:fig3}, at three different energies. The model does not
include the variation of the background with energy, so a flat
background was fitted for each energy independently. In addition,
the overall scattering amplitude had to be systematically reduced
with increasing energy to fit the data satisfactorily. This
reduction, which was nearly a factor of 2 over the energy range
6.5\,meV to 14\,meV, is not predicted by the spin wave model.


The measurements and calculations reported here reveal that the
magnetic correlations in Na$_{0.75}$CoO$_2$ are of a
three-dimensional (3D) nature, despite its highly 2D physical
properties. In fact, the inter-plane exchange constant $J_c$ is
found to be roughly double the intra-plane constant $J_{ab}$. The
spin wave modes propagating along the $c$-axis are found to be
sharp, indicating a well correlated ground state. The in-plane
modes exhibit some broadening, as indicated on Fig.\
\ref{fig:fig4} and reported previously \cite{boothroyd-may04}.

Comparisons have been made between Na$_{x}$CoO$_2$ and other
layered superconducting families, such as the copper oxides. The
strong 2D nature of the cuprates is thought to be important for
their superconductivity, and contrasts with the 3D magnetic
interactions found here for Na$_{0.75}$CoO$_2$. It is likely that
the $c$-axis magnetic coupling is weakened in hydrated
Na$_{x}$CoO$_2$, due to the large inter-layer spacing, and it is
tempting to speculate that this coupling actually inhibits
superconductivity. This possibility is especially pertinent given
recent evidence for the presence of H$_3$O$^+$ ions in the
hydrated compound, which would make the doping level for
superconductivity similar to that in Na$_{0.75}$CoO$_2$
\cite{Argyriou}.

The spin excitation spectrum observed here is not easily
reconciled with the usual picture of localized Co$^{4+}$ and
Co$^{3+}$ ions carrying spins $S=1/2$ and $S=0$, respectively. If
localized Co$^{4+}$ spins were distributed at random then a very
broad magnetic excitation spectrum would be expected, unlike the
sharp modes observed experimentally. One possibility is that there
is a phase separation into ferromagnetic in-plane clusters of
Co$^{4+}$ ions in a matrix of non-magnetic Co$^{3+}$. However, the
Coulomb penalty would be considerable, and to obtain consistency
with the observed sharp spin modes along the $c$-axis these
clusters would have to be aligned vertically above each other over
many layers.

An alternative suggestion based on optical conductivity data
\cite{Bernhard} is that Na$_{0.75}$CoO$_2$ might have a stable
Wigner crystal ground state, with Co$^{4+}$ spins on a triangular
lattice of side $2a$ in a background of Co$^{3+}$. This would
double the period of the magnetic correlations within the layers
and hence create magnetic zone centres at the M-points, e.g.
$(\frac{1}{2}00)$, of the Brillouin zone. Since no magnetic
excitations are observed emerging from the M-points
\cite{boothroyd-may04} we can rule out this type of charge order
in Na$_{0.75}$CoO$_2$, at least in the simplest case where the
Co$^{3+}$ ions are non-magnetic. The Wigner crystal model can only
be reconciled with the observed spin excitation spectrum if the
Co$^{3+}$ ions carry a moment of similar size to the Co$^{4+}$
ions and all moments interact ferromagnetically.


With these points in mind it might be that a more metallic picture
should be sought, in which the Co atoms have no, or only a small
disproportionation of charge. This would naturally produce a
similar magnetic moment on each site, consistent with the model of
an A-type antiferromagnet assumed here. In particular, a weakly
itinerant ground state with strong spin fluctuations could
reconcile the small ordered moment and low ordering temperature
with the relatively large energy scale of the spin excitation
spectrum, and account for the broadening and intensity loss of the
spin modes with increasing energy.

We acknowledge informative discussions with C. Bernhard, P. Bourges, B. Keimer, S. Khaliullin, N. Kovaleva and R. McKenzie. We would like to thank the University of Oxford, the ISIS facility
and the Engineering and Physical Sciences Research Council of
Great Britain for financial support.


\begin{references}

\bibitem{battery-ref}M. M. Doeff {\it et al.}, Electrochem. Acta
{\bf 40}, 2205 (1995) 

\bibitem{thermoelectric-ref} I. Terasaki {\it et al.}, Phys. Rev.
B {\bf 56}, R12685 (1997).

\bibitem{takada-2003} K. Takada {\it et al.}, Nature (London) {\bf 422}, 53
(2003).

\bibitem{structure-ref} R. J. Balsys and R. L. Davis, Solid State
Ionics {\bf 93}, 279 (1996); Q. Huang {\it et al.},
cond-mat/0406570.

\bibitem{Schaak}
R.E. Schaak {\it et al}., Nature {\bf 424}, 527 (2003).

\bibitem{unconventional}
M. Kato {\it et al}., cond-mat/0306036; T. Fujimoto {\it et al}.,
Phys. Rev. Lett. {\bf 92}, 047004 (2004); K. Ishida {\it et al}.,
cond-mat/0308506; Higemoto {\it et al}., cond-mat/0310324; Y.
Kobayashi, M. Yokoi, and M. Sato, J. Phys. Soc. Jpn. {\bf 72},
2161 (2003).

\bibitem{motohashi-2003} T. Motohashi {\it et al.}, Phys. Rev. B
{\bf 67}, 064406 (2003).

\bibitem{susceptibility} D. Prabhakaran {\it et al}.,
cond-mat/0312493; B. C. Sales {\it et al}., cond-mat/0402379.

\bibitem{muSR} J. Sugiyama {\it et al.}, Phys. Rev. B {\bf 67},
214420 (2003); S. P. Bayrakci {\it et al.}, Phys. Rev. B {\bf 69},
100410(R) (2004).

\bibitem{boothroyd-may04} A. T. Boothroyd {\it et al.},
Phys. Rev. Lett. {\bf 92}, 197201 (2004).

\bibitem{In-plane-FM-theory}
D. J. Singh, Phys. Rev. B {\bf 61}, 13397 (2000); D. J. Singh,
Phys. Rev. B {\bf 68}, 020503(R) (2003); G. Baskaran, Phys. Rev.
Lett. {\bf 91}, 097003 (2003); B. Kumar and B. S. Shastry, Phys.
Rev. B {\bf 68}, 104508 (2003); K.-W. Lee, J. Kune\v{s}, and W. E.
Pickett, Phys. Rev. B {\bf 70}, 045104 (2004); M. Mochizuki, Y.
Yanase, and M. Ogata, cond-mat/0407094.

\bibitem{prabhaks-paper} D. Prabhakaran {\it et al.} J. Crystal Growth {\bf 271},
74 (2004).

\bibitem{Chou} F. C. Chou {\it et al.}, cond-mat/0405158.

\bibitem{basis-vectors} The lengths are $|{\bf a}^{\star}| = |{\bf
b}^{\star}| = 4\pi/(a\surd 3)$ and $|{\bf c}^{\star}| = 2\pi/c$,
with $a=2.83$\,{\AA} and $c=10.8$\,{\AA} at room temperature.

\bibitem{Squires}
G.L. Squires, {\it Introduction to the Theory of Thermal Neutron
Scattering} (Cambridge University Press, Cambridge, U.K., 1978).
















\bibitem{Argyriou} C. J. Milne {\it et al.}, cond-mat/0401273

\bibitem{Bernhard}
C. Bernhard {\it et al.}, Phys. Rev. Lett. {\bf 93}, 167003
(2004).

\end{references}
\end{document}